# THIN FILMS OPTICAL PROPERTIES, PHOTOSENSITIVE MATERIALS

## Role of Ge:As ratio in controlling the light-induced response of a-$Ge_xAs_{35-x}Se_{65}$ thin films

Pritam Khan[1], H. Jain[2] and K.V. Adarsh[1]

**Email: pritam@iiserb.ac.in**

[1]*Department of Physics, Indian Institute of Science Education and Research, Bhopal 462023, India.*

[2]*Department of Materials Science and Engineering, Lehigh University, Bethlehem, Pennsylvania 18015, USA*

**In this paper, we present interesting results on the quantification of photodarkening (PD), photobleaching (PB) and transient PD (TPD) in a-$Ge_xAs_{35-x}Se_{65}$ thin films as a function of network rigidity. Composition dependent light-induced responses of these samples indicate that there exist two parallel competing mechanisms of instantaneous PD arising from the As part of the network, and PB arising from the Ge part of the network. Raman spectra of the as-prepared and illuminated samples provide first direct evidence of the light-induced structural changes: an increase in $AsSe_{3/2}$ pyramidal and $GeSe_{4/2}$ corner-sharing tetrahedra units together with new Ge-O bond formation and decrease in energetically unstable edge sharing $GeSe_{4/2}$ tetrahedra. Importantly, for a fixed Se concentration, Ge:As ratio plays the critical role in controlling the net light-induced response rather than the much believed rigidity of the glassy network.**

Unique light-induced optical effects in amorphous chalcogenide thin films make them candidate for potential applications in designing nano-antenna[1], high bit rate waveguides[2] and broadband optical limiters[3]. Light-induced effects in chalcogenide glasses (ChGs) are believed to originate from photogeneration of defect pairs (commonly known as Valence Alternation Pairs (VAPs))[4]. The rearrangement of such defect pairs result in chemical bond redistribution or reorientation, which accounts for many of the light-induced optical and electronic properties[5,6]. Among the many light-induced effects, the most studied effects are photodarkening (PD) in As based ChG[7-10] and photobleaching (PB) in Ge based ChG thin films[6,11]. In a stark contradiction to PD in As-based ChG and PB in Ge based ChG, in our recent work, an unusual coexistence of both PD and PB has been reported in Ge-As-Se ternary compositions at different times[12]. The observed effect is explained by assuming that the non-stochiometric fragments of As-Se and Ge-Se respond to light rather independently.[12] Such a mechanism could not account for the different rates of PD (fast) and PB (very slow). Also its relationship with the network structure or its rigidity, which is most simply described in terms of the mean coordination number (MCN), remained to be established.

Many past studies have shown that network rigidity plays a predominant role in determining the light-induced effects. For example, Calvez *et al.*[13] showed that PD and photoexpansion tend to vanish in rigid region of $Ge_xAs_xSe_{1-2x}$ glass. Notwithstanding, Yang *et al.*[14] and Nemec *et al.*[15] have demonstrated a photostable phase in Ge-As-Se series with very distinct MCN. In this paper, we present the unusual coexistence of PD, PB and transient PD (TPD), and quantification of their magnitude as a function of network rigidity i.e. x in a-$Ge_xAs_{35-x}Se_{65}$ thin films, when illuminated with a 532nm continuous wave (CW) laser. Instantaneous PD dominates the net light-induced response of all samples with $x<10$ and show a crossover to PB when $x>10$. Strikingly, PD and PB do not show a regular trend with respect to MCN; instead they provide the first direct evidence that Ge:As ratio plays a major role in determining the light-induced effects, rather than the much believed rigidity of the glassy network. Raman spectroscopy of these as-prepared and PD/PB samples provides new insights into the light-induced structural rearrangements that establish the coexistence of fast PD and slow PB.

## Results

The X-ray diffraction study on $Ge_xAs_{35-x}Se_{65}$ thin films confirms that they are amorphous in nature (see supplementary Fig. S1 online). The homogeneity and morphology of the film surface are demonstrated by AFM images (see supplementary Fig. S2 online). There is no indication of cracks, defects or sub-micron scale inhomogeneities on the film surface. The surface roughness of all the films is less than 15nm indicating good surface morphology.

In order to study PD/PB, we have recorded transmission spectra of the probe beam through the sample before, during and after pump beam illumination. Fig. 1(a) shows that in the absence of pump beam, there was no change in transmission spectra of the sample, which confirms that the probe beam did not produce any light-induced effects. The data show temporal evolution of transmission ratio ($T_f/T_i$) for all samples at probe wavelengths for which transmission was 20% of the value in the dark or as-prepared condition; the selected probe wavelengths is close to the bandgap of each sample (see supplementary Fig. S3 online). However, in the presence of pump beam, we observe an appreciable shift in $T_f/ T_i$ for all samples. Here $T_i$ and $T_f$ are the transmission of as-prepared sample and transmission after time t since the start of illumination, respectively for the probe wavelength selected as described above. During illumination, $T_f/T_i$ of samples with composition with $x \leq 15$ decreases in a manner that is consistent with PD. Further, the magnitude of PD decreases dramatically and tends towards a photostable regime as x increases from 5 to 15. Then for the film with x=25, $T_f /T_i$ increases with illumination, a clear manifestation of PB. Interestingly, transmission curves of some samples exhibit weak quasi-periodic oscillations with time (e.g. see sample $Ge_{10}As_{25}Se_{65}$ in Fig. 1(a) and $Ge_{15}As_{20}Se_{65}$ in Fig. 1(b)) similar to Abdulhalim oscillation[16-18]. However, their magnitude is too weak in our experiments because of much smaller intensity used compared to the experiment of Abdulhalim *et al.* (more than 1kW/cm$^2$ vs 0.5W/cm$^2$)[18]. At our laser intensity, the material is expected to exhibit predominantly PD and PB; Abdulhalim oscillations are relatively insignificant. The present results clearly demonstrate a crossover from PD to PB in the net light-induced response of a-$Ge_xAs_{35-x}Se_{65}$ thin films, when x exceeds a certain value.

Our in-situ pump-probe experiments provide new insight on the kinetics of the light-induced effects. We note that upon illumination total light-induced changes consist of a transient and a metastable component. When illumination is switched off after complete saturation of PD/ PB, the transient part decays but some switched bonds remain frozen in a metastable state that can only be reversed by annealing near the glass transition temperature ($T_g$)[19]. In the vicinity of $T_g$, the structural groups become sufficiently active so that they can react with each other wihtin the time of observation. However, because of some photolytic reactions, the structure of annealed films is not fully identical to that of the as-prepared films[6]. In this metastable state, indicated by the region between two arrows for a given sample in Fig. 1(a), light-induced effects show dramatically distinct characteristics from the illuminated state described above (see supplementary Fig. S4 online). For the sample with x=5, $T_f/T_i$ increases when we turn off the pump beam, indicating that the sample tries to recover its loss of transparency. However, it settles at a value which is well below the as-prepared state, i.e. sample shows metastable PD. For the sample with x=10, transmission completely recovers to the as-prepared state when we turn off the pump beam, a clear indication that PD is only of transient nature. In contrast to these results, sample with x=15 shows PD in the presence of light initially, but then switches to an overall PB state in the absence of light. At this stage, we envision that there exists a competition between TPD and metastable PB, with TPD dominating in this case. However, when the pump beam is switched off, TPD decays and the sample stabilizes in the PB state. For the sample with x=25, $T_f/T_i$ increases further when we turn off the pump beam, a clear indication that it also possess TPD, however the magnitude is very much lower than PB. Thus our experimental results confirm that a minimum concentration of Ge (in this series x=15) is required to observe PB. Notably, the cross over from PD to PB occurs through a photostable phase. A similar crossover from PD to PB has been observed recently in binary $Ge_xSe_{100-x}$ thin films at ~ x=30 by Kumar *et al.*[20].

Light-induced effects in both the transient and in the metastable regimes call for experiments to quantify PD, TPD and PB. In this context, Fig. 1(a) shows the temporal evolution of the change in transmission ratio ($T_f/T_i$) of all samples for the wavelength at which the initial transmission in the as-prepared state is 20%. From Fig. 1(a) it is evident that upon pump beam illumination, $T_f/T_i$ decreases instantaneously for all samples in a manner which is consistent with PD and eventually saturates within a few tens of seconds. After the complete saturation of PD, $T_f/T_i$ for all samples except x=5 and 10 gradually starts increasing, a clear indication of PB. For the sample with x>15, transmission saturates at a value which is above the as-prepared value (i.e. net PB), and for the samples with x<15, transmission saturates well below the as-prepared value (i.e. net PD). Thus our results demonstrate the coexistence of PD and PB in a-$Ge_xAs_{35-x}Se_{65}$ thin films. To study the transient behaviour, we switched off the pump beam after complete saturation of PD/PB and measured the temporal evolution of $T_f/T_i$ as shown in Fig. 1(a). Quite remarkably, $T_f/T_i$ increases further and saturates very quickly. Repeated on - off cycles clearly indicate that this effect is from TPD, which persists in the sample only during illumination.

After demonstrating the unusual coexistence of PD, TPD and PB in a-$Ge_xAs_{35-x}Se_{65}$ thin films, we have quantified the magnitude of these effects as a function of the structural flexibility of the glassy network. In this context, we define structural flexibility in terms of MCN which is equal to the sum of the respective elemental concentrations times their covalent coordination number[21] and denoted by <r>. Earlier studies showed that the magnitude of light-induced effects decreases with increase in MCN (<r>)[13,22]. A floppy system with lower <r> shows stronger light-induced effect, whereas a rigid system with higher <r> exhibits little to no photostructural changes. The theory predicts that atoms in a floppy media can easily rearrange themselves between on-off states of the laser irradiation, thereby producing large photoeffects. On the contrary, in a rigid system atoms they are relatively constrained, their position remains mostly unaltered throughout the excitation, resulting in weaker light-induced effects. Above observations can be linked to the topography of the energy landscape as well as to the rigidity of the systems having different <r> values[22]. A floppy system is usually associated with high density of minima in the energy landscape and therefore photoexcitation allows the system to explore those minima, thus exhibiting huge light-induced changes. On the contrary, a rigid system having fewer number of configuration states in the energy lanscape undergo weak photostructural changes upon photoexcitation. In this context, PD, TPD and PB are calculated as the difference in transmission between initial (as prepared) and saturated value of PD, difference in transmission between saturated values of on- and off-states of light in the transient regime, and the difference between saturated values of PB and PD, respectively - see Fig. 1(b). Figure 2 shows the variation of PD, PB, TPD and cumulative change with MCN. Interestingly, we found from Fig. 2(a) and 2(b) that PD and TPD do not show a regular trend with MCN, but decrease gradually with increasing MCN. The magnitude of the change in PD and TPD is nearly equal;

here PD is mainly transient and the contribution from metastable PD is small. On the other hand, in Fig. 2(c) PB linearly increases upto a MCN of 2.5 and then shows a sudden jump at 2.6. Notably, from Fig. 2(d) we found that $\Delta T_{cum}$ shows a crossover from PD ($\Delta T_{cum}$ is positive) to PB ($\Delta T_{cum}$ is negative) through a photostable phase at MCN of 2.45. Although $\Delta T_{cum}$ is zero for this sample, we could detect significant TPD without appreciable residual PD/PB. These results clearly demonstrate that PD, TPD and PB do not follow a regular trend as predicted by the network rigidity theory which is already discuued above. It was reported previously that photostable compositions have MCN between 2.45 and 2.55[14, 15, 23] but their transient response remained unknown. A composition that is stable (or metastable) with respect to permanent light-induced changes may still show non-zero transient or reversible changes. Clearly, in situ measurements are required before branding a particular composition fully photostable.

To obtain a more detailed direct information on light-induced structural changes (Ge-Ge, As-As and Se-Se to Ge-Se and As-Se) that are responsible for PD and PB, we have measured the Raman spectra of as-prepared and illuminated samples as shown in Fig. 3. The dominant features of Raman spectra in all the as-prepared films are in the same range of approximately 190 to 270 $cm^{-1}$. They are composed of two independent modes: (1) a sharp peak at 198 $cm^{-1}$ and (2) a broad peak that extends from 224-240 $cm^{-1}$. The Raman peak at 198 $cm^{-1}$ is assigned to the $A_1$ ($v_1$) symmetric vibrational stretching of $GeSe_{4/2}$ corner-sharing tetrahedra. The broad peak from 224-240 $cm^{-1}$ is attributed to the principal vibrational modes of $AsSe_{3/2}$ pyramidal unit and also to minor contributions from $A_1$ ($v_2$) modes of $As_4Se_3$ cage like molecule[24-26]. Apart from these two prominent features, samples with x=10, 15 and 25 show a peak at 215 $cm^{-1}$, which is identified as the companion mode originating from the vibrational edge sharing $GeSe_{4/2}$ tetrahedra[24]. This mode is present only in samples with high Ge concentration. From Fig. 3, it is clear that there is an appreciable change in the intensity of the Raman peaks at 198 and 224-240 $cm^{-1}$ of PD/PB films when compared to the spectra of as-prepared films. We get an insight into the light-induced structural rearrangements if we focus on the edge-sharing $GeSe_{4/2}$ tetrahedra peak at 215 $cm^{-1}$, and the broad Raman mode corresponding to Ge-O at 520-650 $cm^{-1}$ in the illuminated state[27], which are absent in the as-prepared samples. This observation indicates photo-oxidation of Ge in the PB/PD samples. Notably, for PB films, the concentration of edge-sharing $GeSe_{4/2}$ tetrahedra decreases with illumination, a clear indication of light-induced removal of such structures. Our experiments are in agreement with the previous prediction that the local energy in the glass is lower in corner sharing bonds and the laser illumination removes the edge sharing bonds, where the local energy is high[20, 24].

From the Raman spectroscopic study, we find that photo-oxidation of Ge atoms plays a predominant role in producing PB. However, the difference of Raman signal for illuminated and as-prepared sample corresponding to Ge-O Raman mode at 520-650$cm^{-1}$ is small. Therefore, to confirm the oxide formation of Ge atoms, we measured IR absorption spectra of $Ge_{25}As_{10}Se_{65}$ thin films before and after laser illumination. At this point, we refer to Spence *et al.*[28] who also have used difference of IR absorption spectra to detect the chemical changes occuring in the films for Ge-based ChGs. Fig. 4 shows that the difference in IR absorption spectra (ΔA=absorbance after illumination-absorbance in as prepared state) accounts to a broad peak from 540-670$cm^{-1}$, which confirms the occurance of photo-oxidation of Ge atoms in illuminated sample.

**Discussion**

After demonstrating the coexistence of PD, TPD and PB in a-$Ge_xAs_{35-x}Se_{65}$ thin films, it is important to explain the observed effects. PD and PB can be understood by considering the molecular heterogeneities created during thermal evaporation. When the films are illuminated with 532nm light, such compositional heterogeneities associated with As-Se and Ge-Se molecular units respond rather independently. A considerable fraction of metastable homopolar bonds present in the atomic fragments is broken and subsequently converted into energetically favoured heteropolar bonds. The films with MCN<2.45, i.e. with less Ge concentration are over-stoichiometric with respect to As. Upon illumination the As clusters will react in the following way [12, 14]

$$As_{(x-y)}Se_z + yAs + h\nu \leftrightarrow As_xSe_z \tag{1}$$

and give rise to PD. The net PB in films having MCN >2.45 i.e. with less As content can be explained by assuming light-induced structural rearrangement at non-stochiometric Ge sites and the creation of Ge-Se bond at the expense of Ge-Ge / Se-Se bonds by the following reaction [12, 29]

$$Ge-Ge+Se-Se+h\nu \leftrightarrow 2Ge-Se \tag{2}$$

The Ge-Se bond is stronger compared to Ge-Ge and Se-Se bond. Therefore, it is favoured as the sample attempts to reach equilibrium with minimum possible free energy. Another plausible mechanism for the observed PB is the photo-oxdiation of Ge atoms by the creation of Ge-O bonds at the expense of Ge-Ge bonds that are broken by illumination to form $GeO_4$ structural units[30, 31]. The ab initio density functional theory calculations for $GeSe_2$ suggest that the formation of Ge–Ge bond has no significant effect on band gap, but only on the broadening of the Ge 4s band[32, 33]. On the other hand, Se–Se bonds significantly decrease the bandgap because of the increase in the highest occupied molecular orbit (HOMO) states in the valence band formed by Se(p) non-bonding lone-pair (LP) molecular orbital, and at the same time with no appreciable change in the lowest unoccupied molecular orbit (LUMO) of the conduction band formed of σ*antibonding molecular orbital of Ge(s)–Se(p)[32]. Furthermore, time dependent density functional theory has also demonstrated that there is a significant reduction in optical bandgap if the glassy network has edge sharing $GeSe_{4/2}$ tetrahedra[32]. In this context, we have analyzed the Raman data of all the samples in more detail and calculated the ratio of the peak intensity of the Raman mode corresponding to corner sharing $GeSe_{4/2}$ tetrahedra (Ge-CS) and $AsSe_{3/2}$ pyramidal units (As-P) - see Table 1. The results show that with illumination the ratio of Ge-CS/As-P decreases for x=15 and 25, whose photo response is dominated by PB. For example, for the sample with x=25, the value of Ge-CS/As-P ratio in the as-prepared state is 2.28, which is reduced to 1.89 in the PB state. This observation shows that As-Se bond density increases and Ge-Se bond density decreases with illumination. Apart from this, we observed the Raman mode at 520-650 $cm^{-1}$ corresponding to Ge-O in illuminated sample, which was absent in the as-prepared samples (also confirmed by IR spectroscopy – see Fig.4). This Raman mode confirms photo-oxidation of Ge in the PB samples. The Raman mode at ~215 $cm^{-1}$ corresponding to edge sharing $GeSe_{4/2}$ tetrahedra, is decreased by light illumination, which also strongly favors PB (also supported by DFT calculations by Holomb *et al.*[32] on $GeSe_2$ glass). From the above analysis, we conclude that there exist two parallel light-induced processes: PB due to photo-oxidation of Ge and decrease in edge sharing $GeSe_{4/2}$ tetrahedra, and PD from the increase in $AsSe_{3/2}$ pyramidal units. Photo-oxidation is a relatively slower process than the other two. The same conclusion was reached by Yan *et al.*[34] from observations on $xGe_{45}Se_{55}$-(1-x)$As_{45}Se_{55}$. As a result, if we look at the temporal evolution of the transmission spectra, we observe an initial fast PD followed by a relatively slower PB as observed from Fig. 1. Thus the present experimental results establish a direct relationship between the light-induced structural rearrangement and photo-oxidation on the coexistence of fast PD and a slow PB. Transient PD (TPD) can be thought to be originating from the light-induced bond switching and atom movement, similar to that in As-based ChGs. By comparison, TPD can be understood by assuming the apperarnce of an intermediate state of electron transitions between the ground and photoexcited state[29].

To model the reaction kinetics the two opposite photoeffects, we use a combination of stretched exponential functions that describe PD and PB separately:

$$\Delta T = C[\exp\{-( t/\tau_d )^{\beta_d}\}] + \Delta T_{Sd} + \Delta T_{Sb}[1-\exp\{-( t/\tau_b )^{\beta_b}\} \qquad (3)$$

where the subscripts 'd' and 'b' correspond to PD and PB, respectively. $\Delta T_S$, τ, β and t are metastable part, effective time constant, dispersion parameter and illumination time, respectively, and C is a temperature dependent quantity which is equal to the maximum transient changes. The net rate equation for the whole process is a summation of respective PD and PB. The experimental data fit very well to eq. (3) - see Fig. 5. Fitting parameters calculated from theoretical fit are listed in Table 2. From the Table 2, it is evident that the effective reaction time for PD is relatively short, a few tens of seconds. By contrast, PB is a slower process compared to PD, with much longer reaction times. Notably, for the sample with x=10, we get a different picture from the above descreiption for x=15 and 25. First, this sample shows the presence of edge sharing $GeSe_{4/2}$ tetrahedra in as-prepared as well as in illuminated state, indicating such configuration is relatively stable to light illumination. Secondly, the ratio of Ge-CS/As-P remains unchanged at 1.18 for both as-prepared and illuminated states. Thirdly, there is only a small increase in the intensity of Ge-O bonds. Therefore, we envision that above mentioned effects together are responsible for the observed photostability in the metstable regime. In the case of sample with x=5, the absence of Ge-O and edge sharing $GeSe_{4/2}$ tetrahedra (in both as-prepared and illuminated states) together with the increase in $AsSe_{3/2}$ pyramidal units give rise to an effective PD and the complete absence of PB. Thus we obtain direct evidence of light-induced structural rearrangement in a-$Ge_xAs_{35-x}Se_{65}$ thin films and demonstrate that such effects can be effectively controlled by choosing an optimum composition.

In summary, we report an unusual coexistence of PD, TPD and PB in a-$Ge_xAs_{35-x}Se_{65}$ thin films and quantified the magnitude of these effects as a function of the structural flexibility of the glassy network. There is a composition dependent competition between two parallel mechanisms viz. rapid PD and a slower PB. The former dominates for samples with x<10 via changes in As-Se bonds, and the latter dominates for compositions with x>10 through photo-oxidation of Ge together with the decrease in edge sharing $GeSe_{4/2}$ tetrahedra. The sample with x=10 shows photostability in the metastable regime, mainly because of the stability of edge sharing $GeSe_{4/2}$ tetrahedra, as seen in similar value of Ge-CS/As-P ratio both in as-prepared and in illuminated state together with resistance to oxidation of Ge atoms. Further, for a fixed Se (chalcogen) concentration, PD and PB do not show a simple correlation with MCN i.e. rigidity of the glassy network; instead Ge:As ratio determines the net light-induced response.

**Methods**

**Sample Preparation.** Four bulk $Ge_xAs_{35-x}Se_{65}$ glasses with x = 5, 10, 15 and 25 were prepared starting with 99.999% pure Ge, As and Se powders and using the melt-quench method .The cast samples were used as the source material for depositing thin films of average thickness ~1.0 μm by thermal evaporation in a vacuum of about $1 \times 10^{-6}$ Torr.

**Pump-probe spectroscopy.** PD and PB were studied by using an in situ pump probe optical absorption method described previously[35]. In our experiment, a diode pump solid state laser (DPSSL) of wavelength 532 nm with an intensity of $0.5 W/cm^2$ was used as the pump beam and the probe beam was a low intensity white light. Changes in the transmission of the probe beam were recorded in the wavelength range 470-850 nm and also at certain selected wavelengths close to the optical bandgap of the sample at a time interval of 225 ms.

**IR Spectroscopy.** IR spectra (4000–400 cm-1) were recorded by coating the films on KBr substrate and was recorded using a Perkin Elmer Spectrum BX spectrophotometer.

**Morphological and Structural Characterization.** To study the amorphous nature of thin films, they were characterized by X-ray diffraction with Cu Kα radiation. AFM (Agilent Tech., Model 5500) images were recorded in contact mode using DPE-18 cantilever (tip diameter 10 nm with 75 Hz frequency and 3.5 N/m force constant). Raman spectra of as-prepared and illuminated films were obtained with Horiba Lab RAM high resolution spectrometer using the 632.8 nm excitation light from a He-Ne laser (power = 15 mW). Since He-Ne laser can induce some light-induced effects, we avoided it carefully by using low intensity excitation beams and also making sure that its illumination on the sample lasted for a short duration of about 20 seconds.

**Acknowledgments**

**The authors thank Department of Science and Technology (Project no: SR/S2/LOP-003/2010) and council of Scientific and Industrial Research, India, (grant No. 03(1250)/12/EMR-II) for financial support. They gratefully acknowledge the US National Science Foundation for supporting international collaboration through International Materials Institute for New Functionality in Glass (DMR-0844014). We would like to thank Amit Adhikary for IR measurements. We also thank Ananya Patra for some scientific discussions.**

**Author contributions**

K. V. A conceived the idea. P. K. made the samples and did the experiments. P. K., H. J. and K. V. A. analyzed the data and wrote the manuscript.

**Additional information**

Competing financial interests: The authors declare no competing financial interests.

**Table 1 | Intensity ratio of corner sharing $GeSe_{4/2}$ tetrahedra (198cm$^{-1}$) and $AsSe_{3/2}$ pyramidal units (230-240 cm$^{-1}$) in as-prepared and illuminated state.**

| | $GeSe_{4/2}$(CST)/$AsSe_{3/2}$(P) | |
|---|---|---|
| Sample | As-prepared state | Illuminated state |
| $Ge_5As_{30}Se_{65}$ | 0.80 | 0.79 |
| $Ge_{10}As_{25}Se_{65}$ | 1.18 | 1.18 |
| $Ge_{15}As_{20}Se_{65}$ | 1.33 | 1.28 |
| $Ge_{25}As_{10}Se_{65}$ | 2.28 | 1.89 |

**Table 2 | Time constants obtained from Eq. 1 that corresponds to PD and PB for a- $Ge_xAs_{35-x}Se_{65}$ films. The subscript d and b refer to darkening and bleaching, respectively.**

| Sample | $\tau_d$ (sec) | $\beta_d$ | $\Delta T_{Sd}$ | $\tau_b$ (sec) | $\beta_b$ | $\Delta T_{Sb}$ |
|---|---|---|---|---|---|---|
| $Ge_5As_{30}Se_{65}$ | 25 | 0.45 | 0.87 | 0 | 0 | 0 |
| $Ge_{10}As_{25}Se_{65}$ | 25 | 0.65 | 0.84 | 0 | 0 | 0 |
| $Ge_{15}As_{20}Se_{65}$ | 26 | 0.65 | 0.90 | 980 | 0.65 | 0.05 |
| $Ge_{25}As_{10}Se_{65}$ | 38 | 0.58 | 0.94 | 1896 | 0.80 | 0.28 |

**Figure 1 |(a) Temporal evolution of $T_f$ /$T_i$ of all a-$Ge_xAs_{35-x}Se_{65}$ thin films for the wavelength at which transmission is 20% of the value for the dark condition. Dashed line indicates the time at which laser was turned on. The downward (wine color) and upward (pink color) arrows represent the time at which laser is turned off and on respectively, showing the transient effects. (b) Schematic diagram showing the calculation of PD, TPD, PB and cumulative change in relative transmission for a-$Ge_{15}As_{20}Se_{65}$.**

**Figure 2 | Quantification of magnitude of PD, PB, TPD and cumulative effect as a function of MCN of the network. MCN increases from 2.4 to 2.6 when x increases from 5 to 25 in a-$Ge_xAs_{35-x}Se_{65}$ thin films.**

**Figure 3 | Raman Spectra of as-prepared and illuminated (a)a-$Ge_5As_{30}Se_{65}$ (b) a-$Ge_{10}As_{25}Se_{65}$ (c) a-$Ge_{15}As_{20}Se_{65}$ and (d) a-$Ge_{25}As_{10}Se_{65}$ thin films.**

**Figure 4 | IR difference absorption spectra (after-before illumination) for a-$Ge_{25}As_{10}Se_{65}$ thin films.**

**Figure 5 | Time evolution of $T_f$/$T_i$ for a- $Ge_{25}As_{10}Se_{65}$ thin film for the wavelength at which the as prepared transmission is 20%. The blue hollow circles and red lines represent the experimental data and theoretical fit, respectively.**

**Fig.1**

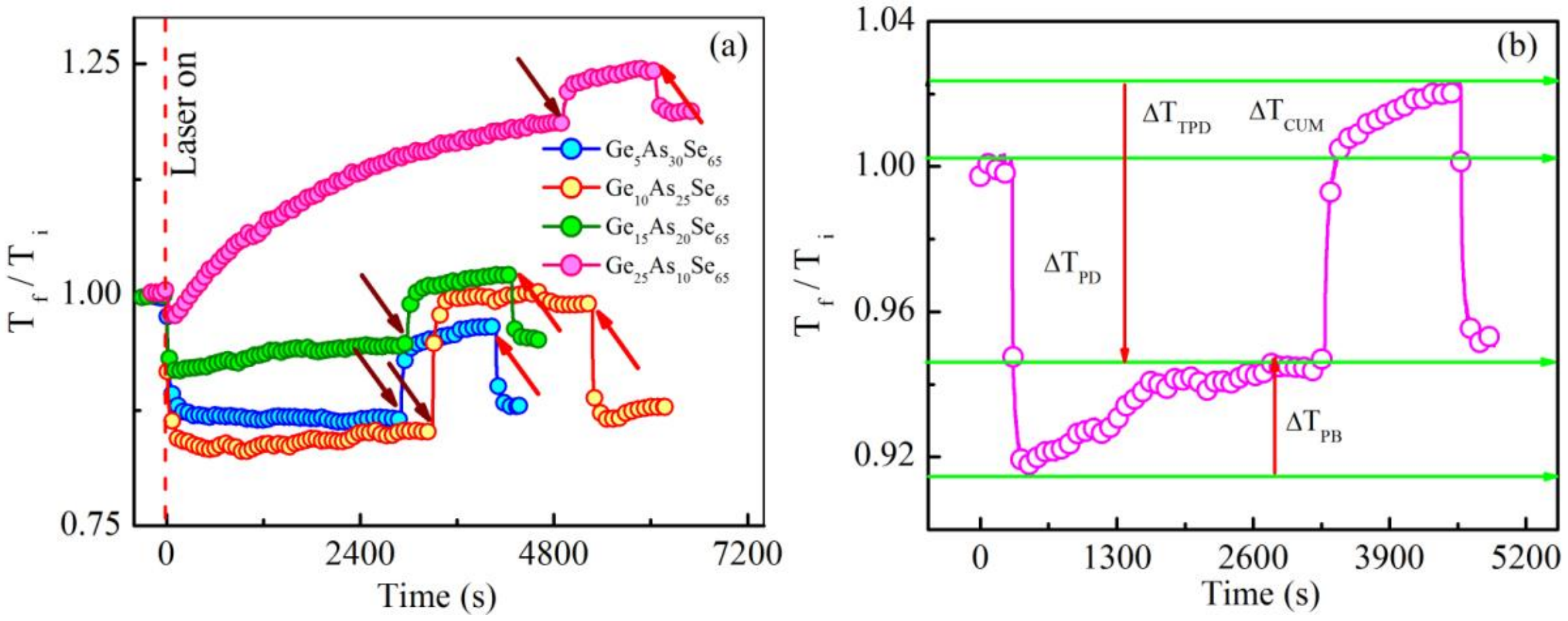

**Fig. 2**

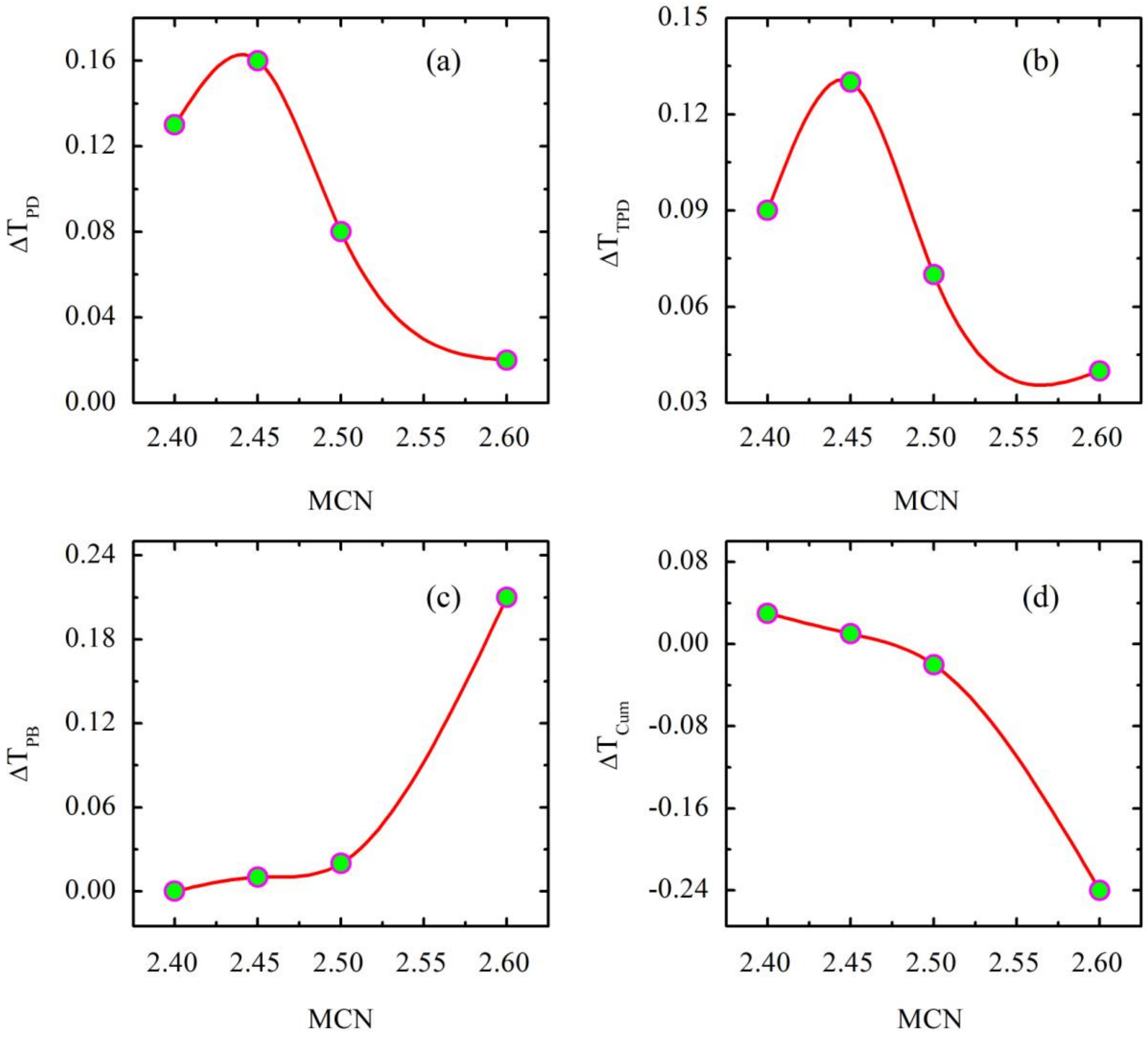

**Fig. 3**

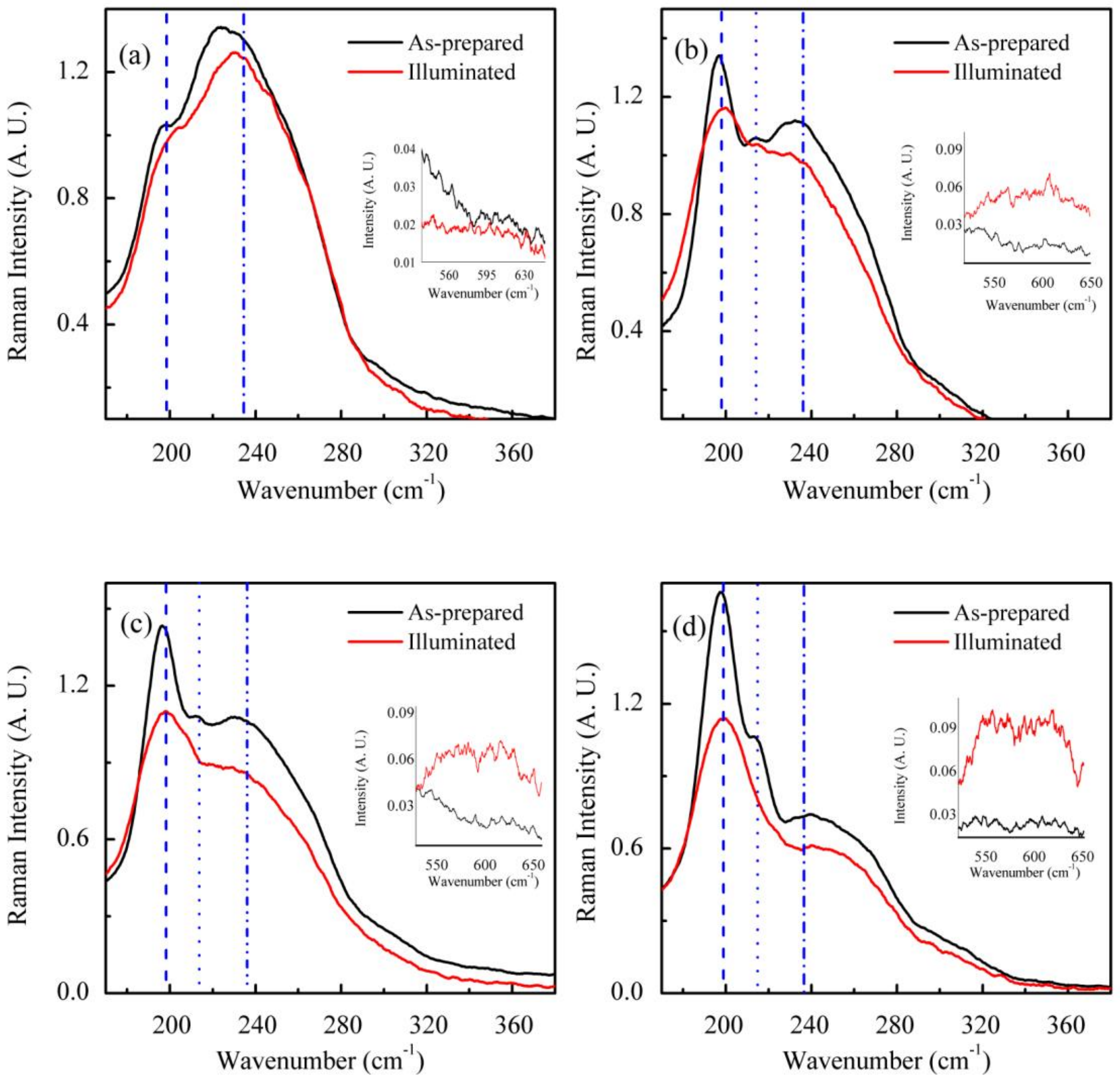

**Fig. 4**

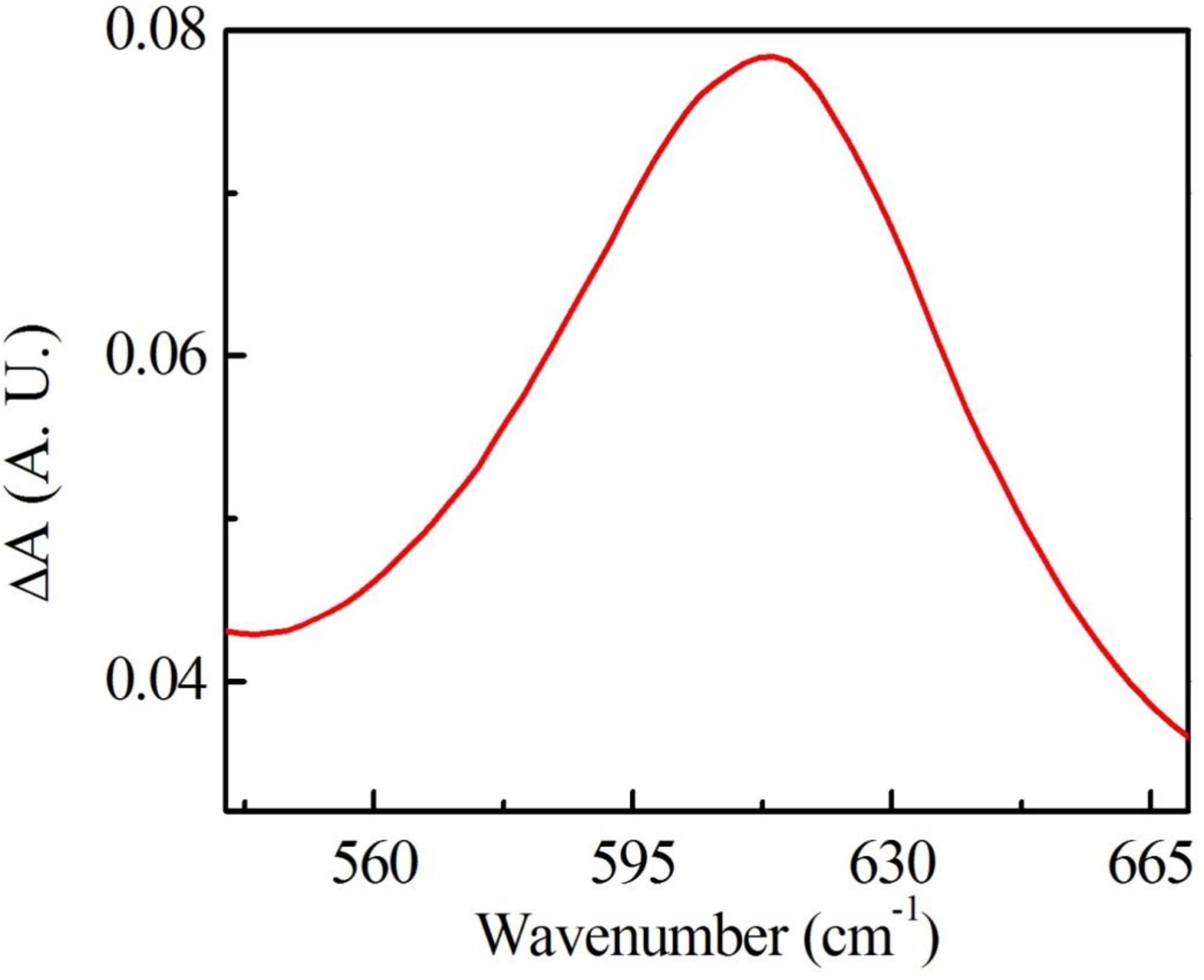

**Fig. 5**

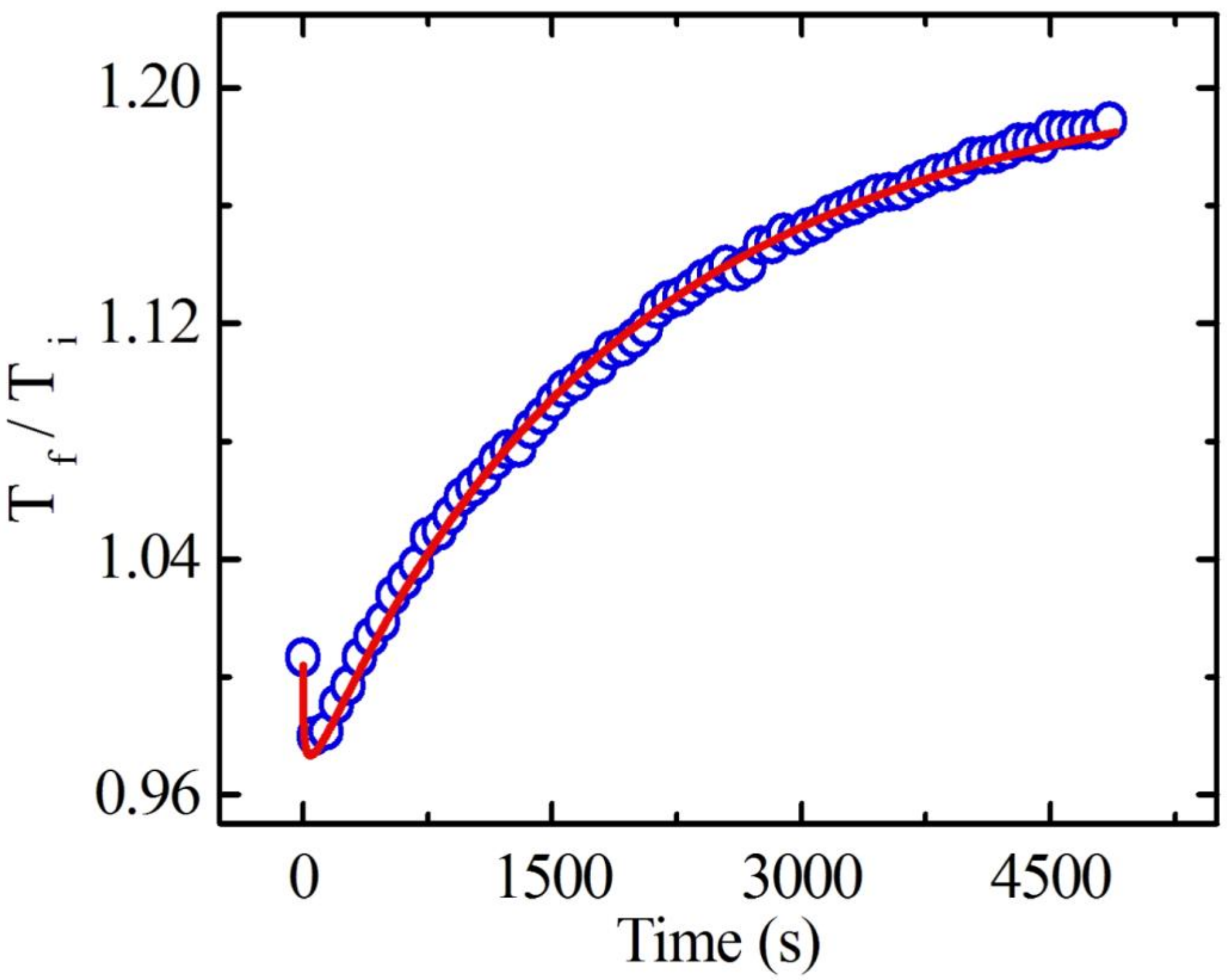

## Supporting Information

## Role of Ge:As ratio in controlling the light-induced response of a-$Ge_xAs_{35-x}Se_{65}$ thin films

Pritam Khan[1], H. Jain[2] and K.V. Adarsh[1*]

*[1]Department of Physics, Indian Institute of Science Education and Research, Bhopal 462023, India.*

*[2]Department of Materials Science and Engineering, Lehigh University, Bethlehem, Pennsylvania 18015, USA*

## Supplementary figures

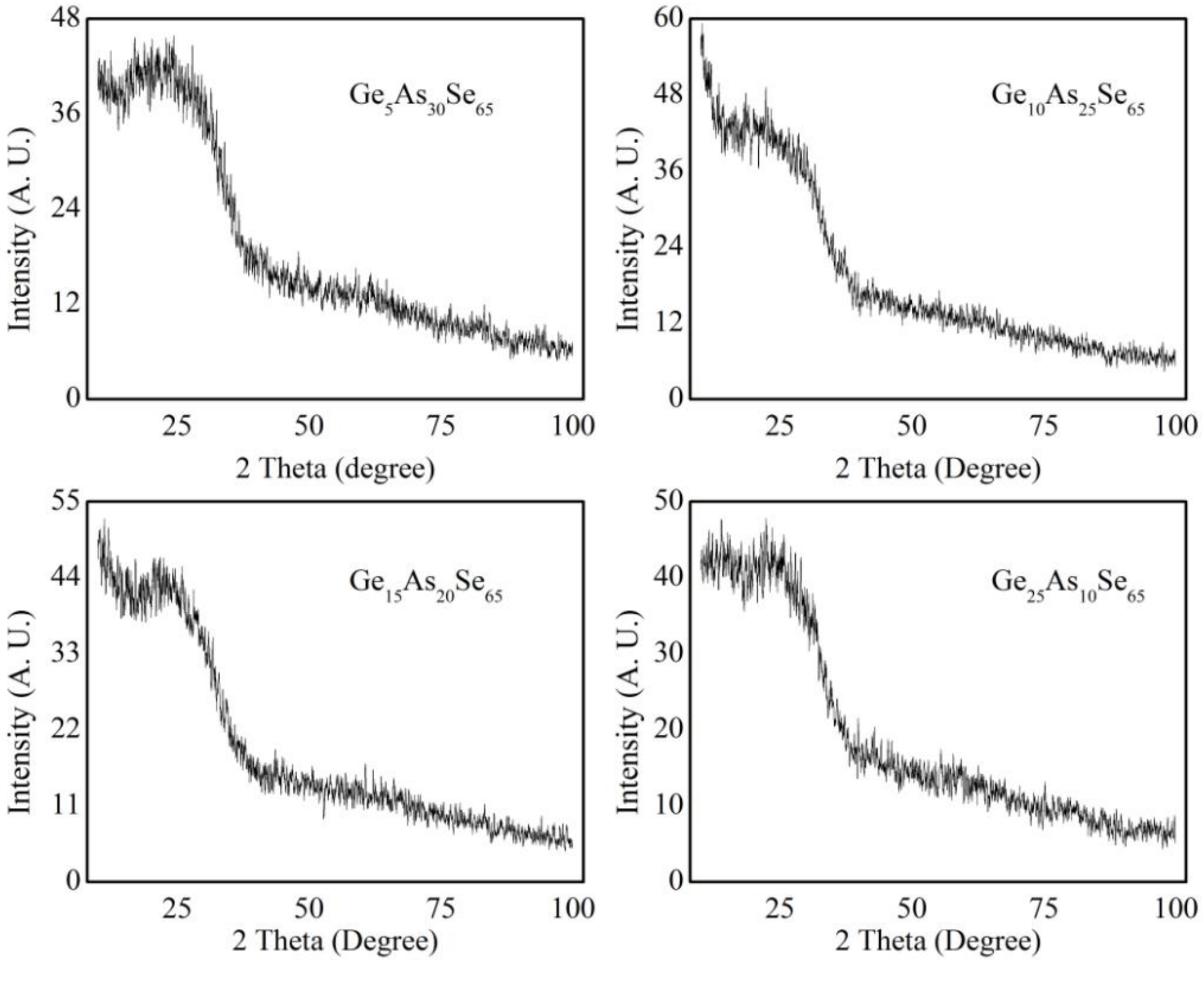

***Figure S1.*** XRD images of $Ge_xAs_{35-x}Se_{65}$ thin films. It is quite evident from the figure that films are amorphous in nature.

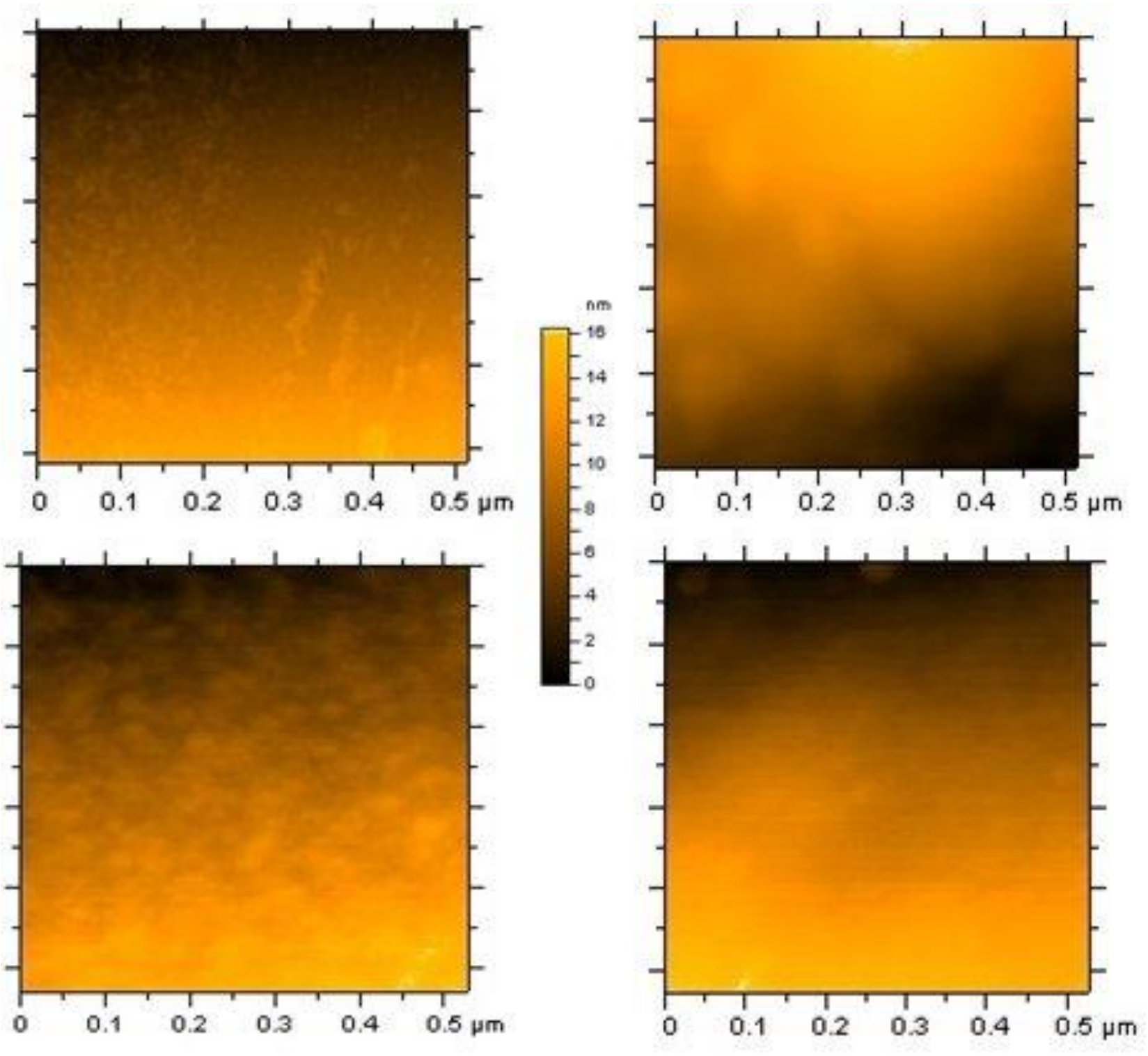

***Figure S2.*** AFM images of $Ge_xAs_{35-x}Se_{65}$ thin films. Figures clearly indicate that film surface is homogeneous with no cracks or defects.

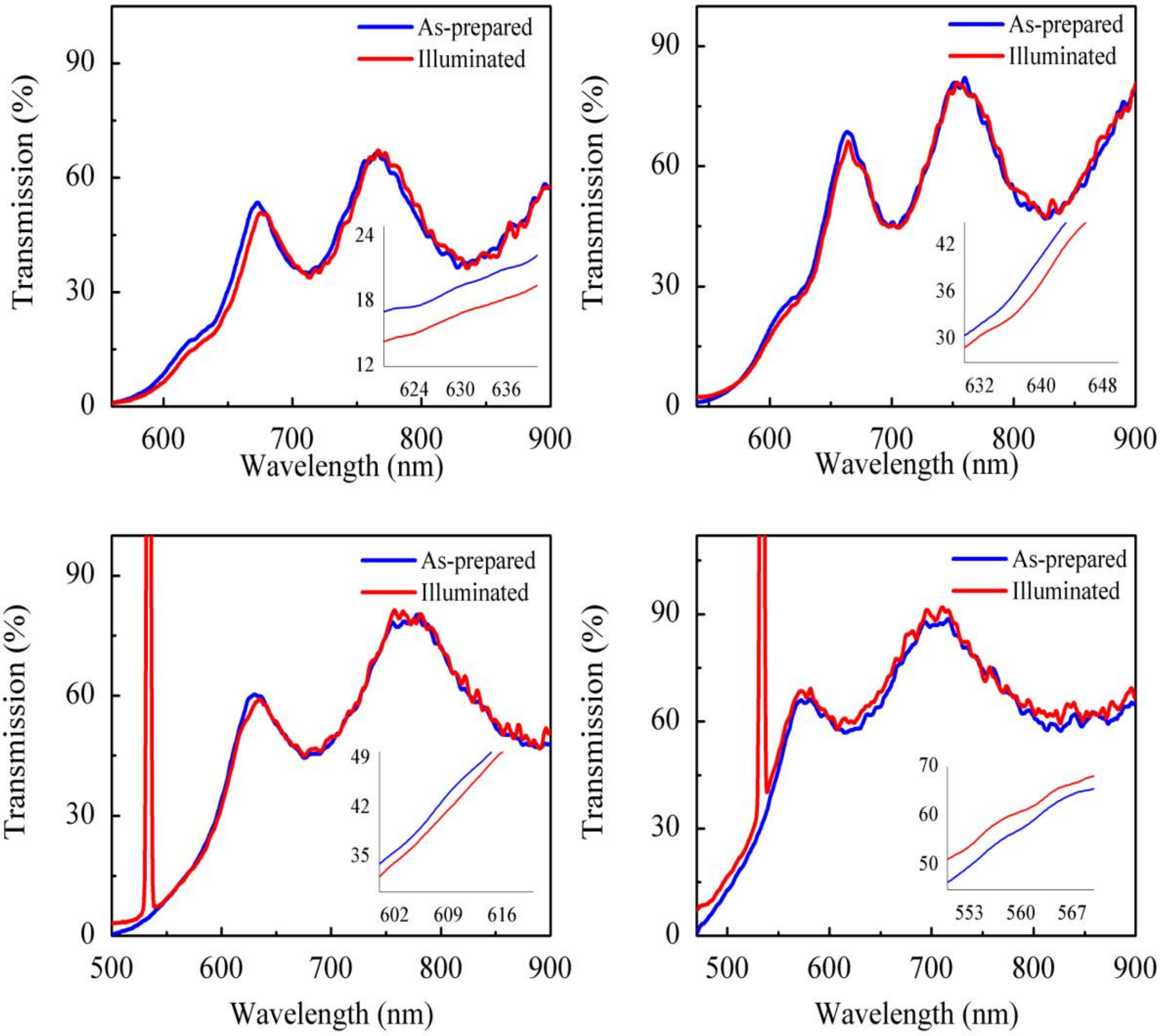

***Figure S3***. Transmission spectra of (a) a-$Ge_5As_{30}Se_{65}$ (b) a-$Ge_{10}As_{25}Se_{65}$ (c) a-$Ge_{15}As_{20}Se_{65}$ (d) a-$Ge_{25}As_{10}Se_{65}$ thin films in as-prepared and illuminated state. Inset shows expanded region near to bandgap.

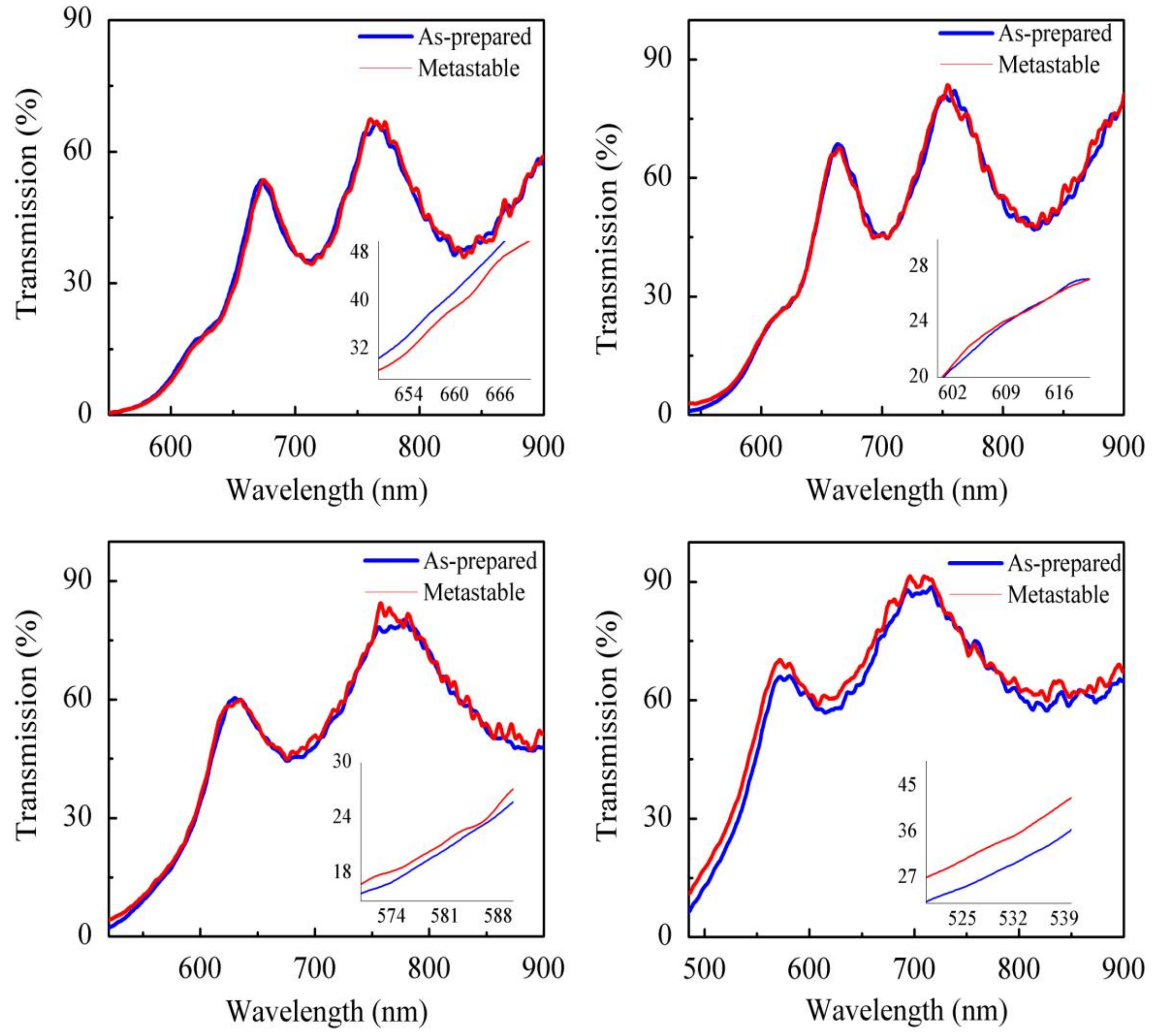

***Figure S4***. Transmission spectra of (a) a-$Ge_5As_{30}Se_{65}$ (b) a-$Ge_{10}As_{25}Se_{65}$ (c) a-$Ge_{15}As_{20}Se_{65}$ (d) a-$Ge_{25}As_{10}Se_{65}$ thin films in as-prepared and metastable state (after turning off the pump beam). Inset shows expanded region near to bandgap.